\documentclass[useAMS,usenatbib]{mn2e}
\usepackage[dvips]{graphicx}
\usepackage{longtable}
\usepackage{setspace}
\def\p1{\phi^{\prime}}

\def\otriangle{\mbox{$\triangle\hspace{-0.072 in}\cdot$\hspace{0.072 in}}}

\begin{document}

\title[Spatial and Kinematical Lopsidedness] {Spatial and Kinematical 
Lopsidedness of Atomic Hydrogen in the Ursa Major Group of Galaxies}
 
\author[R.A.Angiras et al.]{R.A.Angiras$^{1}${\thanks {On leave from St. Joseph's College, Bangalore,India}}, C.J.Jog $^2$,K.S.Dwarakanath$^3$,M.A.W. Verheijen$^4$\\
1.School of Pure and Applied Physics,M.G. University, Kottayam, 686560,India \\
2.Department of Physics, Indian Institute of Science, Bangalore, 560012, India \\
3.Raman Research Institute, Bangalore, 560080, India\\
4.Kapteyn Astronomical Institute, Rijksuniversiteit Groningen,Netherlands\\
(E-mails:rangiras{@}rri.res.in,cjjog{@}physics.iisc.ernet.in,dwaraka{@}rri.res.in,verheyen{@}astro.rug.nl)}

\date{Accepted.....; Received .....}


\maketitle


\begin{abstract} 

We have carried out the harmonic analysis of the atomic hydrogen (HI) 
surface density maps and the velocity fields for 11 galaxies belonging 
to the Ursa Major group, over a radial range of 4-6 disc scalelengths in each galaxy. This analysis gives the radial variation of
spatial lopsidedness, quantified by the Fourier amplitude A$_1$ of the m=1 component normalised to the average value. The kinematical
analysis gives a value for the elongation of the potential to be
 $\sim 10 \% $. The mean amplitude of spatial lopsidedness is found to be $\sim 0.14$ in the inner disc, similar to the field galaxies, and is smaller by a factor of $\sim 2$ compared to the Eridanus group galaxies. It is also shown that the the average value of A$_1$ does not increase with the Hubble type, contrary to what is seen in field galaxies.  
We argue that the physical origin of lopsidedness in the Ursa Major group of galaxies is tidal interactions, albeit weaker and less frequent
than in Eridanus. Thus systematic studies of lopsidedness in
groups of galaxies can provide dynamical clues regarding the  
interactions and evolution of galaxies in a group environment.

\end{abstract}
 
 
\section{INTRODUCTION}

It is well-known that the distribution of the  
atomic hydrogen gas (HI) in many spiral galaxies is non-axisymmetric or lopsided. This lopsidedness was first studied by Baldwin, Lynden-Bell, \& Sancisi (1980).
From the analysis of global profiles of HI using single dish radio 
telescopes, it was concluded that over 50 \% of galaxies studied show 
such asymmetry (Richter \& Sancisi 1994, Haynes et al. 1998, 
Matthews, van Driel, \& Gallagher 1998). Due to the lack of
spatial resolution, such studies could only reveal the combined effect of spatial and velocity asymmetry.
The lopsidedness has also been observed kinematically, as in terms of 
asymmetric rotation curves (Swaters et al. 1999) which can be
directly used to obtain the lopsided perturbation potential
(Jog 2002). The kinematical asymmetry has been used to deduce the
 ellipticity of the potential
from the analysis of the HI velocity fields (Franx, van Gorkom, \& de Zeeuw 
1994; Schoenmakers et al. 1997). 

Many theories have been proposed for the observed lopsidedness. They include 
tidal interactions (Jog 1997), minor mergers \citep{Zaritsky97}, asymmetric gas accretion \citep{Bournaud05}, 
and offset stellar disc in the halo potential \citep{Noord01}. It is not yet clear which of these possible scenarios work at scales of a galactic group. 

Recently, in a first such study, the 2-D spatial distribution of HI was 
Fourier-analysed to measure the lopsidedness for a sample of 18 galaxies in the 
Eridanus group of galaxies (Angiras et al. 2006, here onwards called Paper I). 
In this paper, it was found that the mean amplitude of measured lopsidedness 
is large $\sim 0.2$ (i.e., a surface density contrast of 20\% above an uniform disc), nearly twice of that seen in the stellar component 
of the field galaxies (Rix \& Zaritsky 1995, Bournaud et al. 2005).
Also, it was shown that the early-type spiral galaxies 
show a higher lopsidedness than the late-type spirals, which is contrary
to that observed in field galaxies (Bournaud et al. 2005). These two results suggest 
that tidal interactions play a dominant role in the generation of lopsidedness in 
the galaxies in groups, as was argued in paper I.
It is not known whether these results are common to all group
environments, especially since the groups are known to exhibit a
large variety of properties \citep{Rasmussen06}, especially regarding the galaxy density and velocity dispersions.

Here we address this issue by analysing the 2-D HI data for 
galaxies in the Ursa Major group. The velocity dispersion, the fraction of early type galaxies and the lack of HI deficiency \citep{Marc01,Omar05} are the major differences between the Ursa Major and the Eridanus groups.
Thus the galaxies in the Ursa Major group provide an opportunity to study lopsidedness in a different physical environment. We have selected 11 spiral galaxies belonging to the Ursa Major Group for estimating the spatial 
lopsidedness and elongation.  The spatial lopsidedness values are compared with the results of the Eridanus group of galaxies (Paper I) to see whether group environments give rise to similar values for the amplitudes and phases of the spatial asymmetries. The higher order Fourier components (m=1,2,3) have also been obtained. Further, we have carried out an analysis of the kinematical data to estimate the elongation in the potential. This is compared with  the values obtained from the spatial analysis.As in the Eridanus case (Paper I), the use of HI as a tracer allows us to study lopsidedness to outer disks covering a radial distance larger than twice that studied earlier using the near-IR data on stars as in Rix \& Zaritsky (1995) and Bournaud et al. (2005).

This paper is organised as follows. In section 2 we discuss the HI and optical data used for the proposed analysis. Details of the
harmonic analysis, and the results are presented in section 3. A discussion of these results in the context of the group environment is given in section 4 and in section 5 we give the conclusions.

\section{Data}
\subsection{The Ursa Major Group}
The Ursa Major group in the super galactic co-ordinate system lies between $58.53\le SGL \le 73.53$ degrees and
$-4.46 \le SGB \le 10.54$ degrees \cite{Tully96}. The systemic velocity of this group is $950$ km s$^{-1}$ with a
dispersion of $150 $km s$^{-1}$ \cite{Marc01}. Seventy nine galaxies have been associated with this group \cite{Tully96} and this group had a projected density of 3 galaxies per Mpc$^{-2}$. Due to the small velocity dispersion, it is not yet clear whether it can be classified as a cluster as mentioned in Tully et al.(1996). In addition this
group contains a smaller fraction of early-type galaxies ($\sim 14\%$ (E+S0) and a larger fraction of late-type 
galaxies $\sim 86\%$ (Sp+Irr)\cite{Tully96}) This system does 
not show a central concentration \cite{Marc01} which is atypical for a cluster and more similar to that found in a group.

As we shall be using the results from a similar analysis carried out on Eridanus group by Angiras et al. (2006) 
in section 3\&4, a brief summary of the characteristics of this group is given. This group in super-galactic 
co-ordinate system lies between $\sim -30\le SGL \le -52$ degrees and $272\ge SGB \ge 292$ degrees at 
a mean distance of $\sim 23\pm2$Mpc. Approximately 200 galaxies are associated with this group with a velocity 
dispersion of $\sim 240$kms$^{-1}$. In this case the projected density was higher, $\sim$ 8 galaxies per
 Mpc$^{-2}$ \citep{Omar05}. 
Unlike Ursa Major, in Eridanus group there was a larger fraction of elliptical and S0 type galaxies ($\sim 30\%$ (E+S0) and a smaller fraction of the late-type galaxies $\sim 70\%$ (Sp+Irr) \citep{Omar05}). In this case also, like Ursa Major, the group centre is not known. The selection criteria for the 18 galaxies that were used for the harmonic analysis by Angiras et al. (2006) are similar to what is given in section 2.2 (paragraph 1). 

Even though, these two groups are almost at the same distance from us they differ mainly in two aspects. Firstly, in Eridanus group, HI deficiency is seen which is ascribed to tidal interactions \citep{Omar05b}. HI deficiency is not seen in the case of Ursa Major \citep{Marc01}. Secondly, compared to Eridanus, Ursa Major is a loose group \citep{Tully96,Omar05}.

The higher number density and the higher velocity dispersion as seen in the Eridanus group implies a higher rate of tidal interactions between the galaxies in the group. Also, the
higher fraction of early-type galaxies seen in the Eridanus represent an earlier evolution of galaxies via tidal 
interactions. These agree with the higher amplitude of lopsidedness seen in the Eridanus 
galaxies, if generated by tidal interactions.
We caution, however, that the spatial distribution of the Ursa Major galaxies is that of 
an elongated filament \citep{Tully96}. Hence, it may not be so straightforward to calculate the dependence of the galaxy 
interaction rate on the number density in that case.

\subsection{Radio Data}

Out of the 49 galaxies observed using the Westerbork Synthesis Radio Telescope (WSRT) by Verheijen and Sancisi (2001) , we have
selected 11 galaxies on the basis of their inclination and quality of HI maps (galaxies, whose HI maps were patchy were rejected) for further analysis. The right ascension ($\alpha$),
declination ($\delta$), systemic velocity ($V_{sys}$), inclination ($i$) and Position Angle (PA) of these galaxies are
given in Table 1. All the galaxies selected were in the inclination range of 45 to 70 degrees.  This was to ensure the availability of good
resolution in velocity maps and HI maps, both of which were essential for the analysis \citep{Block02,Bournaud05}. Details of the
observation and the preliminary data reduction are given elsewhere \cite{Marc01}.

\begin{table*}
\centering
\noindent
\caption{The sample of galaxies selected for spatial lopsidedness analysis \citep{Marc01}}
\begin{tabular}{@{}lcrrccc@{}}
\hline
\hline
\bf{Name}     &Hubble Type& \bf{$\alpha$}\small(J2000) & \bf{$\delta$}\small(J2000) &\bf{$V_{sys}$}&Inclination&Position Angle\\
              &	& ~h~~m~~s~ & ~~\hbox{$^\circ$}~~$'$~~$''$~& (km s$^{-1}$)&($^\circ$)&($^\circ$) \\
                                                                                                                             
\hline
UGC 6446      &Sd&11~26~40.4& 53~44~48&644.3&54&200\\
NGC 3726      &SBc&11~33~21.2& 47~01~45&865.6&54&194\\
NGC 3893      &Sc&11~48~38.2& 48~42~39&967.2&49&352\\
NGC 3949      &Sbc&11~53~41.4& 47~51~32&800.2&54&297\\
NGC 3953      &SBbc&11~53~48.9& 52~19~36&1052.3&62&13\\
UGC 6917      &SBd&11~56~28.8& 50~25~42&910.7&59&123\\
NGC 3992      &SBbc&11~57~36.0& 53~22~28&1048.2&58&248\\
UGC 6983      &SBcd&11~59~09.3& 52~42~27&1081.9&50&270\\
NGC 4051      &SBbc&12~03~09.6& 44~31~53&700.3&50&311\\
NGC 4088      &Sbc&12~05~34.2& 50~32~21&756.7&71&231\\
NGC 4389      &SBbc&12~25~35.1& 45~41~05&718.4&50&276\\
\hline
\hline\\
\end{tabular}
\end{table*}

For the sake of completeness, a brief summary of the data reduction procedure is given here. As a result of observation
of typical duration of 12 to 60 hour with WSRT, raw UV data were obtained. These data were calibrated, interactively
flagged and Fast Fourier Transformed using the NEWSTAR software. The resulting data cubes were further processed using
the Groningen Image Processing SYstem (GIPSY). All the data cubes were smoothed to $30^{\prime\prime}\times
30^{\prime\prime}$ and continuum subtraction was carried out. The resulting cubes were used to derive the HI-surface
density (Moment 0) and HI-velocity (Moment 1) maps. The typical 3$\sigma$ column density of $10^{20}$cm$^{-2}$ was obtained for the moment 0 maps.
The moment 1 maps had typical velocity resolution of $\sim 19$kms$^{-1}$. It should be emphasised that the Eridanus angular resolution of $20^{\prime\prime}$ ($\sim 2.24$kpc) and velocity resolution ($\sim 10$kms) and column density \citep{Omar05} were comparable to that of Ursa Major.

\subsection{Optical and Near-IR Data}

The K$^\prime$-Band and R-Band images of a few of the largest galaxies having a typical diameter of $3^{\prime}
-6^{\prime}$ in the inclination range of $49^\circ - 62^\circ$ were sourced from the Canadian Astronomy Data
Centre (CADC)  archives. These images were
obtained using various telescopes and CCD cameras by Tully et al. (1996) and is kept in the archives after the initial data
reductions like cosmic ray removal, dark subtraction, and flat fielding were carried out.
The typical
resolution of the images were $~1^{\prime\prime}$ (R-Band) and $\sim 2^{\prime\prime}$ (K-Band). These are analysed to obtain the asymmetry in the stellar 
distribution, and compare 
that with the HI asymmetry (Section 3.2).

\section{Harmonic Analysis}

\subsection{Harmonic Analysis of Radio Data}

We have adopted the harmonic analysis for analysing the data (Paper I). In HI,
where both velocity maps and surface density maps are available, the 
analysis technique is different from that adopted
in optical analysis. The procedure assumes that in an ideal galaxy, HI is in pure circular motion. Hence we have,

\begin{equation}
V(x,y)=V_{0}+V_{c}\cos(\p1)\sin(i)+V_{r}\sin(\p1)\sin(i)
\end{equation} 

\noindent where $V(x,y)$ is the velocity at the rectangular coordinate $(x,y)$,$V_{0}$ is the systemic velocity, $V_{c}$ is the rotation velocity, $i$ is the
inclination and $V_{r}$ is the expansion velocity which was taken to be zero. The azimuthal angle ($\p1$) measured in the plane
of the galaxy, is given by the equations

\begin{equation}
\cos(\p1) ={\frac {-(x-x_{0})\sin(PA)+(y-y_{0})\cos(PA)}{r}}
\end{equation}
\begin{equation}
\sin(\p1)={\frac{-(x-x_{0})\cos(PA)+(y-y_{0})\sin(PA)}{r\cos(i)}}
\end{equation}

\noindent where $r=\sqrt{((x-x_{0})^2+(y-y_{0})^2/\cos(i)^2)}$. In these equations, $(x_{0},y_{0})$ is the kinematical centre of
the galaxy, $PA$ is the position angle of the galaxy measured in the anti-clockwise direction from the north direction.
Using these equations, the five unknown parameters, namely $(x_{0},y_{0})$, PA, $V_c$ and $i$ were estimated using the GIPSY
task ROTCUR \citep{Baldwin80} in an iterative manner \citep{Wong03,Omar05}. It was observed that the dynamical centre, derived from
velocity maps were less than $2^{\prime\prime}$ away from the optical centre. 
Hence, for all the calculations the optical centre was used. 

\subsubsection{Spatial Lopsidedness and Other Non-axisymmetry in HI}

The harmonic coefficients were derived from the surface density maps, assuming that the surface density at
each radii can be expanded in the form

\begin{equation}
I(r,\p1)= a_0(r) + \sum_{m}a_m\cos m[\p1 - \phi_{m}(r)]
\end{equation}

Here, $a_m$ is the amplitude of the surface density harmonic coefficient and $\phi_{m}(r)$ is the phase. The harmonic coefficients so
derived were normalised using the mean surface density ($a_0$) at each radius. The variation of 
the normalised amplitude of the first order harmonic
coefficient A$_1 (=a_1/a_0)$ and of the phase angle $\phi_1$ with respect to 
the radius are shown (Figures 1 \& 2).The values for the average A$_1$ measured in the larger range 1.5-2.5 $R_{K'}$ is given in column 4 of table 2. These values can be compared with the values estimated by earlier workers \citep{Angiras06,Bournaud05,Rix95}. Similarly the values of the fractional Fourier
amplitudes A$_1$, A$_2$, and A$_3$ 
corresponding respectively to the Fourier components m=1,2,3 in the range 1-2 R$_w$are given in Table 2 (columns 6,7 \&)- see Section 3.3 .

\begin{figure*}
\includegraphics[width=84mm,height=25mm]{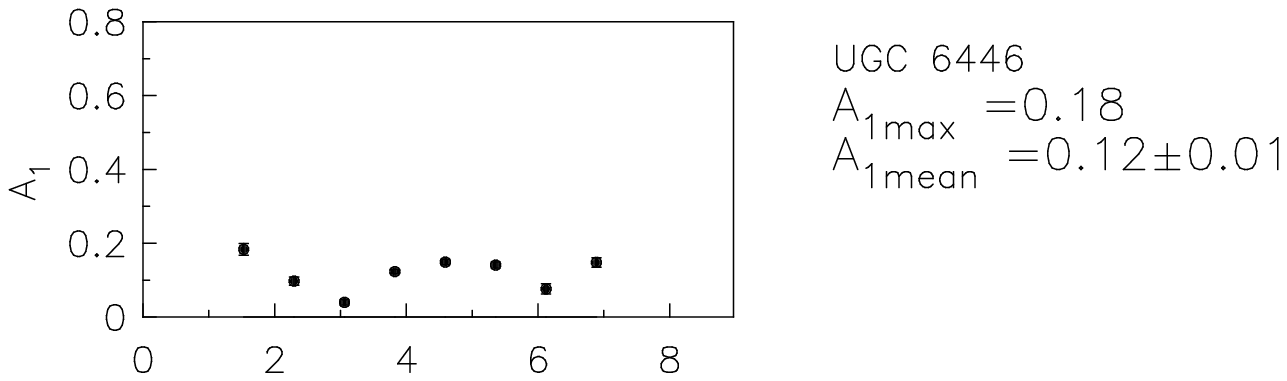}
\includegraphics[width=84mm,height=25mm]{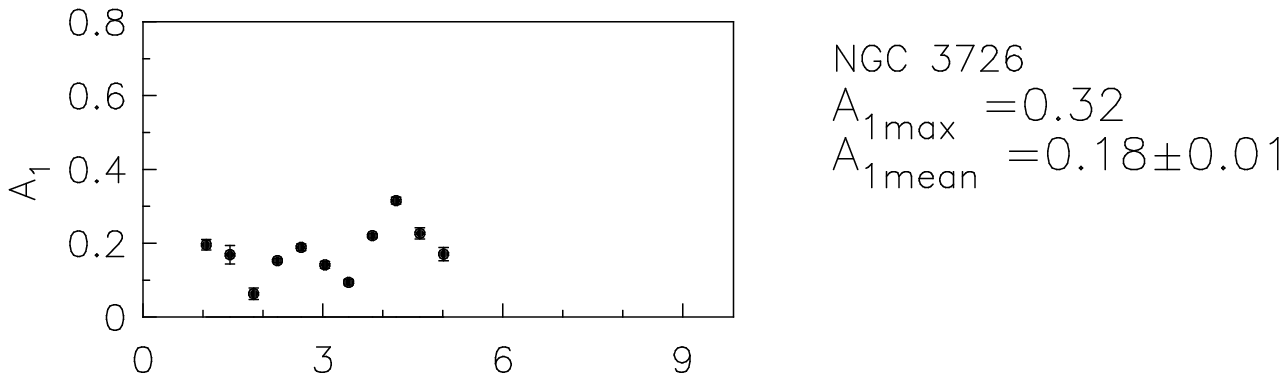}
\includegraphics[width=84mm,height=25mm]{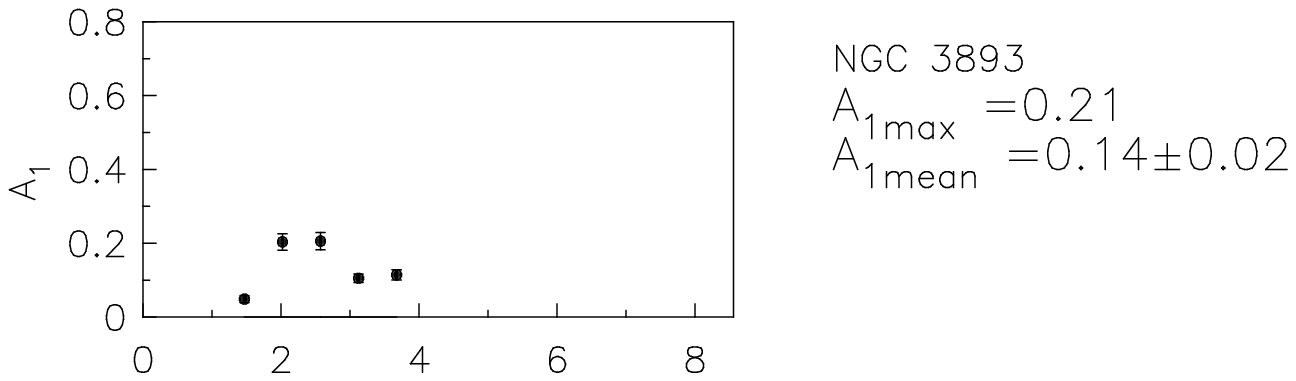}
\includegraphics[width=84mm,height=25mm]{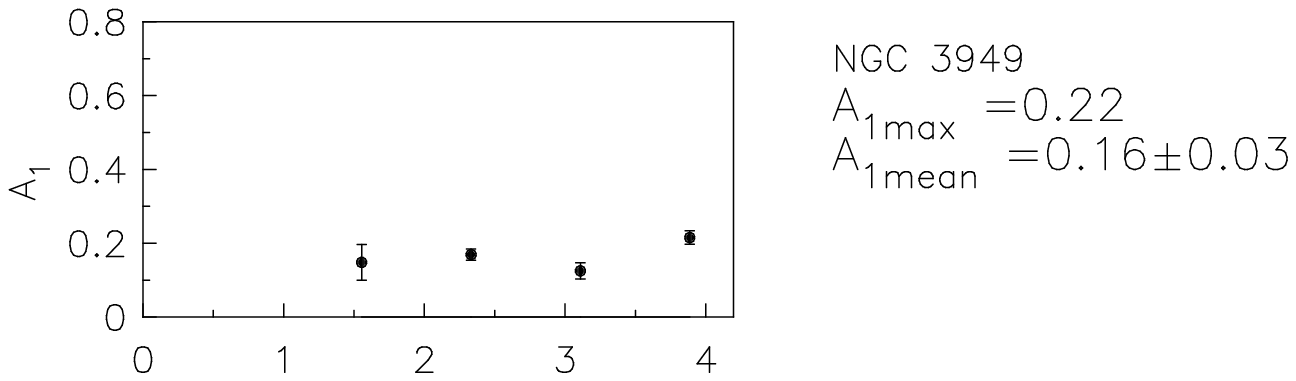}
\includegraphics[width=84mm,height=25mm]{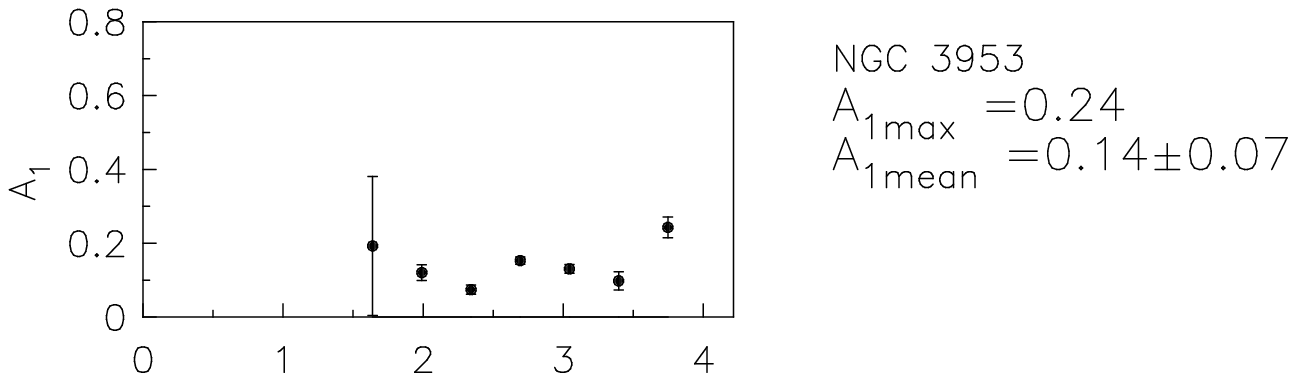}
\includegraphics[width=84mm,height=25mm]{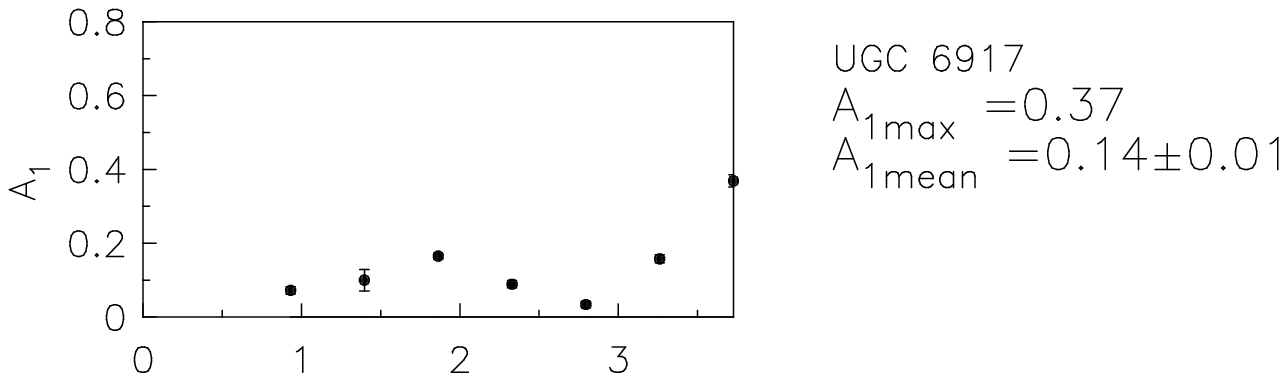}
\includegraphics[width=84mm,height=25mm]{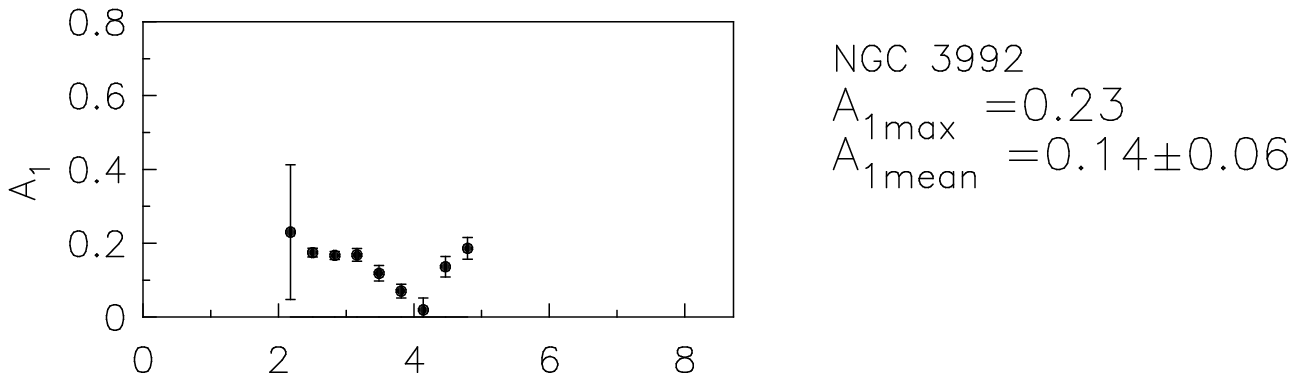}
\includegraphics[width=84mm,height=25mm]{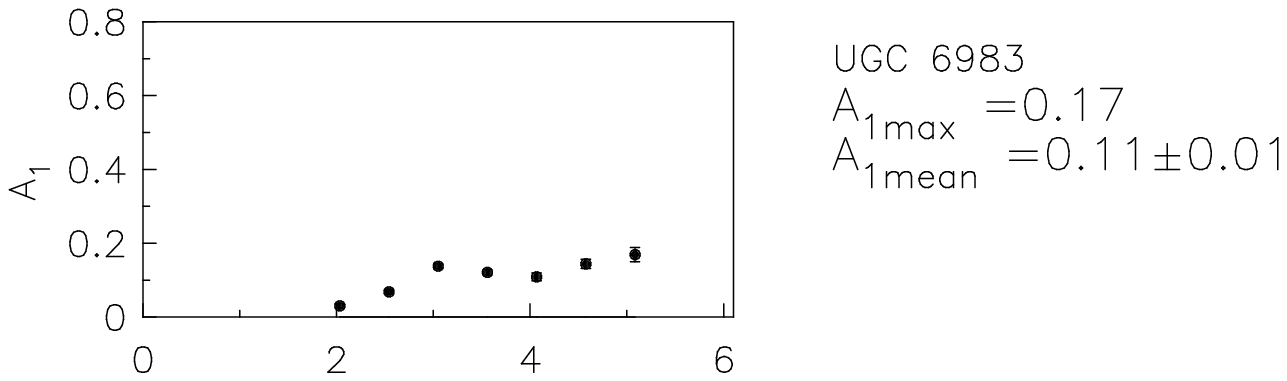}
\includegraphics[width=84mm,height=25mm]{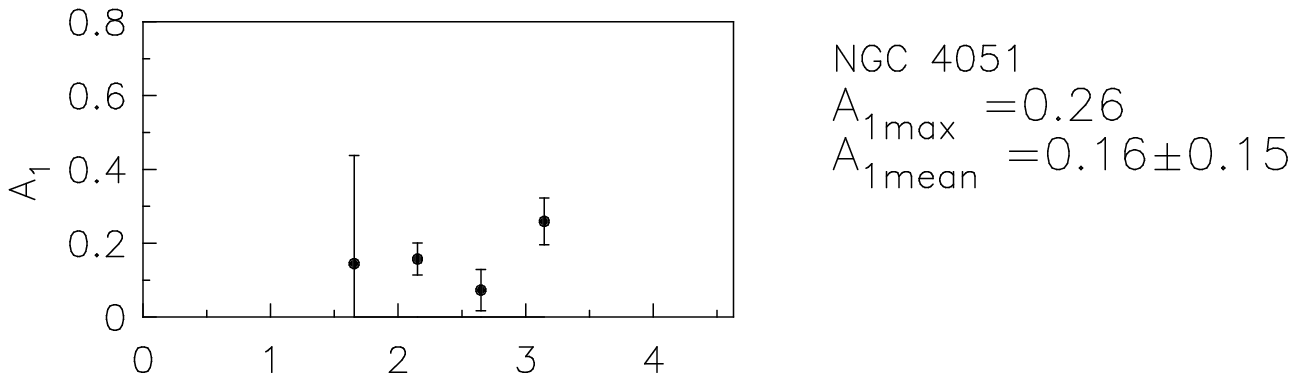}
\includegraphics[width=84mm,height=25mm]{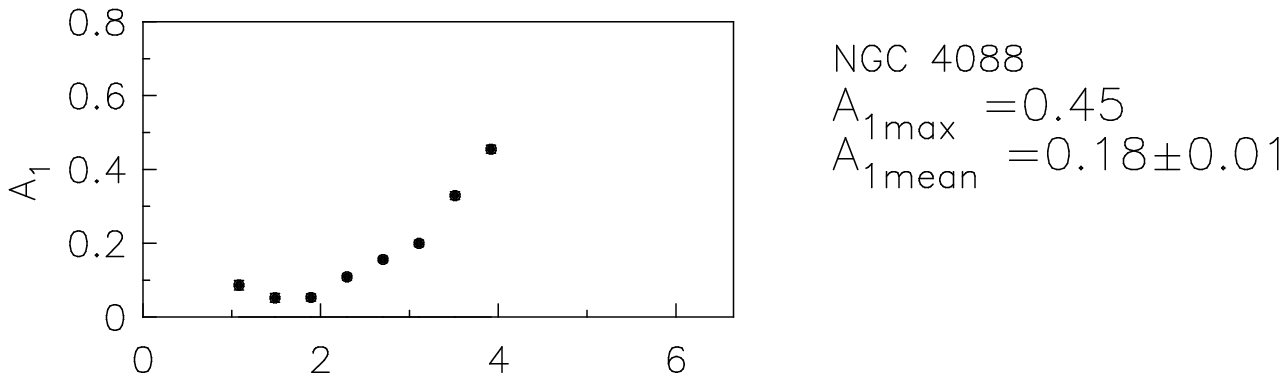}
\includegraphics[width=84mm,height=25mm]{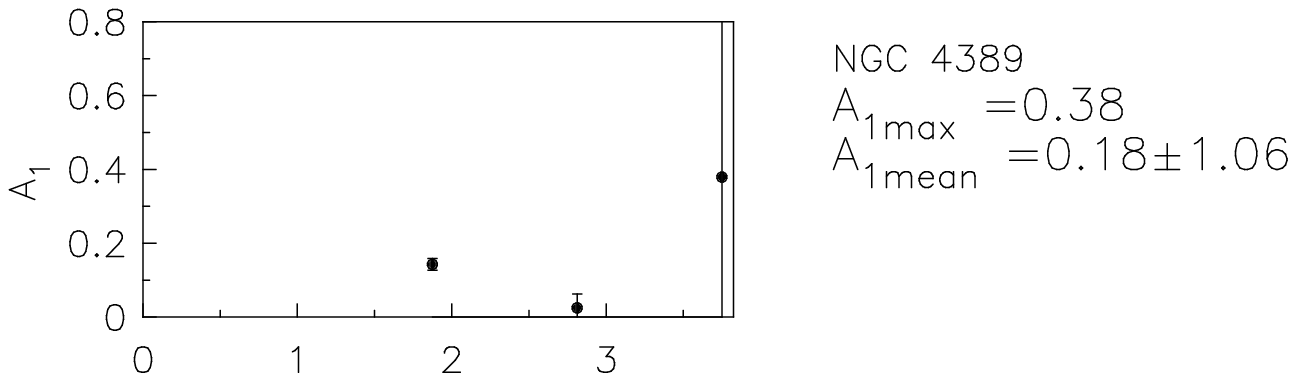}
\caption{ The asymmetry parameter derived from the surface density maps (moment 0). In each of the maps, the radius is in the units of K'-band scale length. The mean value estimated is for the complete range.}

\end{figure*}

\begin{figure*}
\includegraphics[width=84mm,height=25mm]{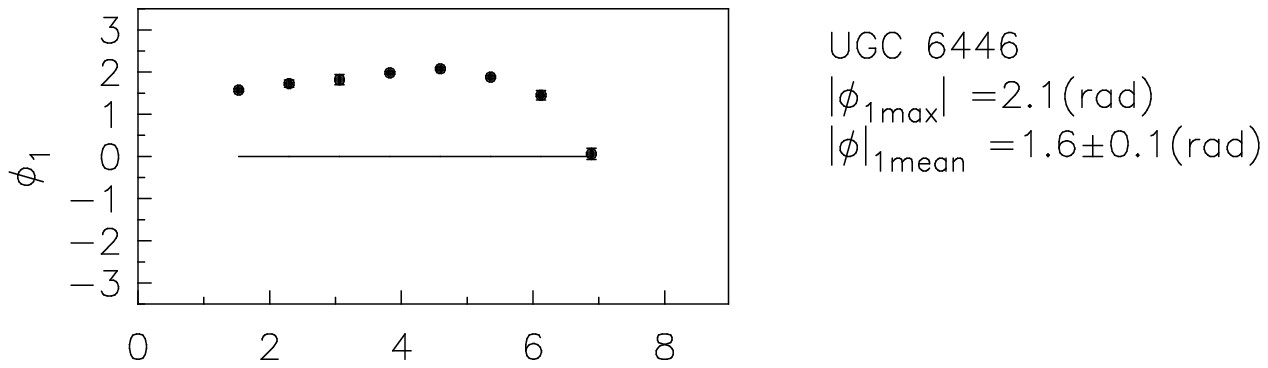}
\includegraphics[width=84mm,height=25mm]{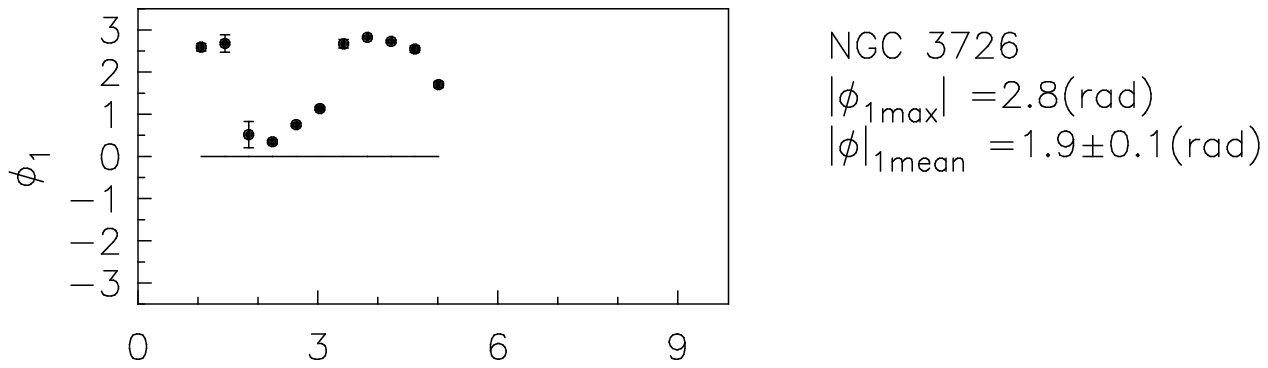}
\includegraphics[width=84mm,height=25mm]{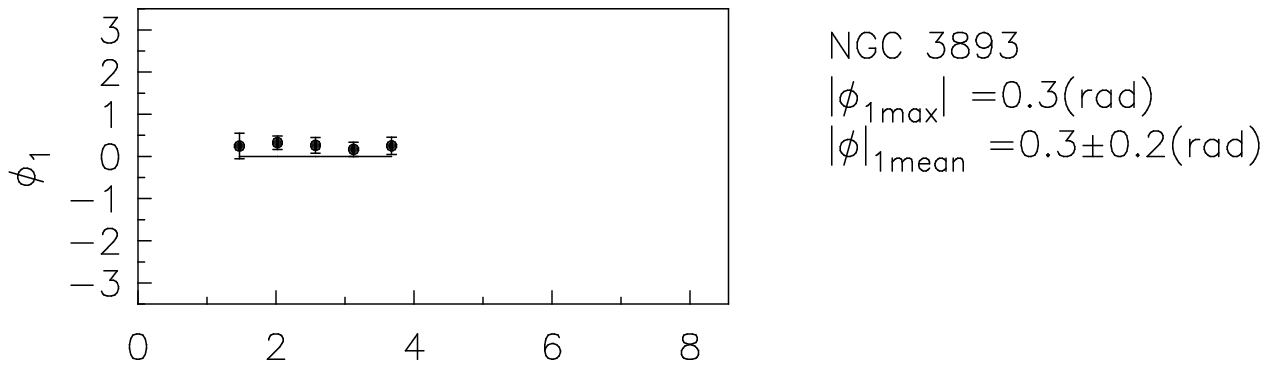}
\includegraphics[width=84mm,height=25mm]{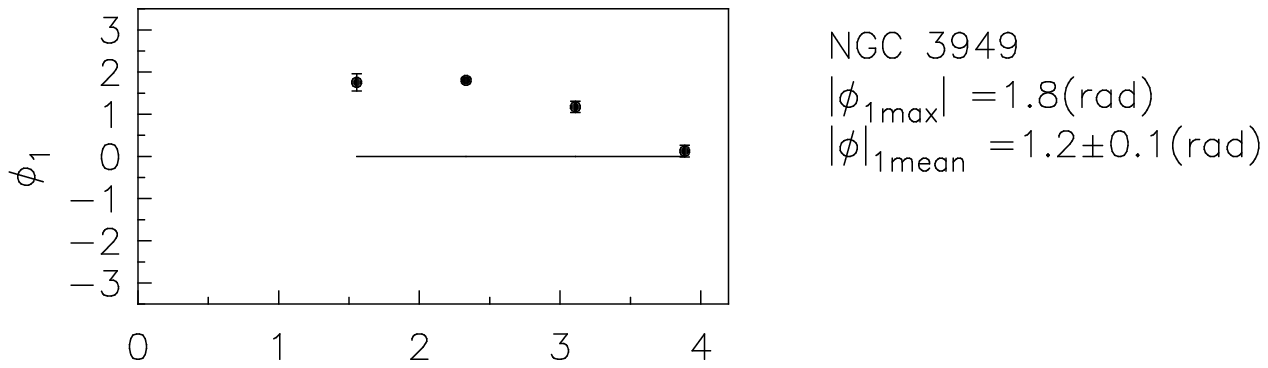}
\includegraphics[width=84mm,height=25mm]{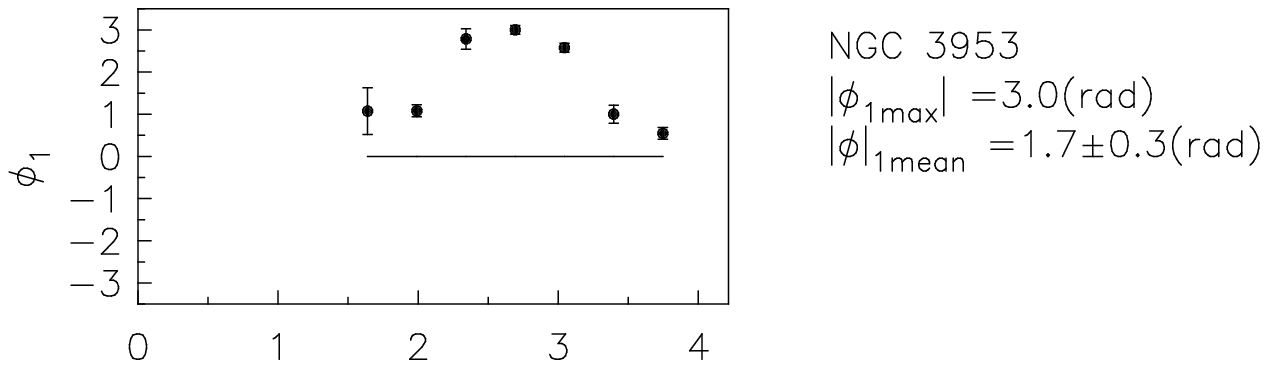}
\includegraphics[width=84mm,height=25mm]{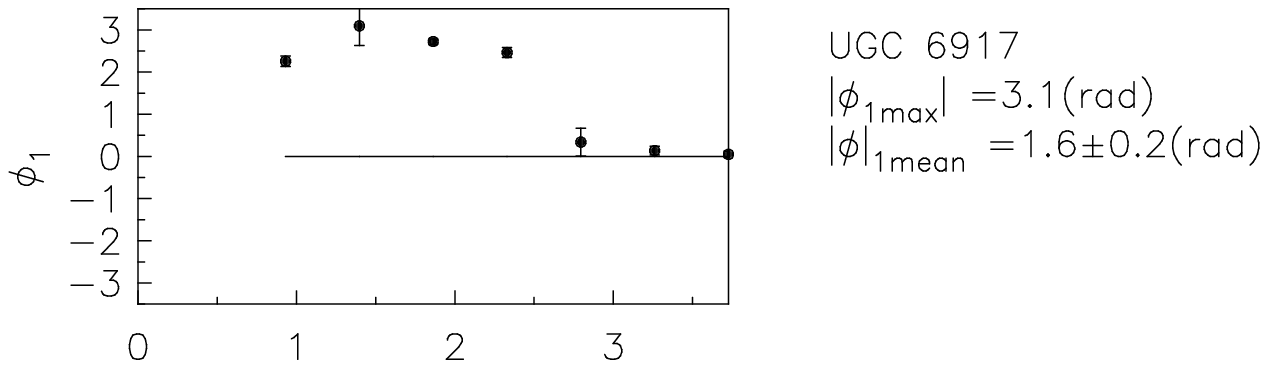}
\includegraphics[width=84mm,height=25mm]{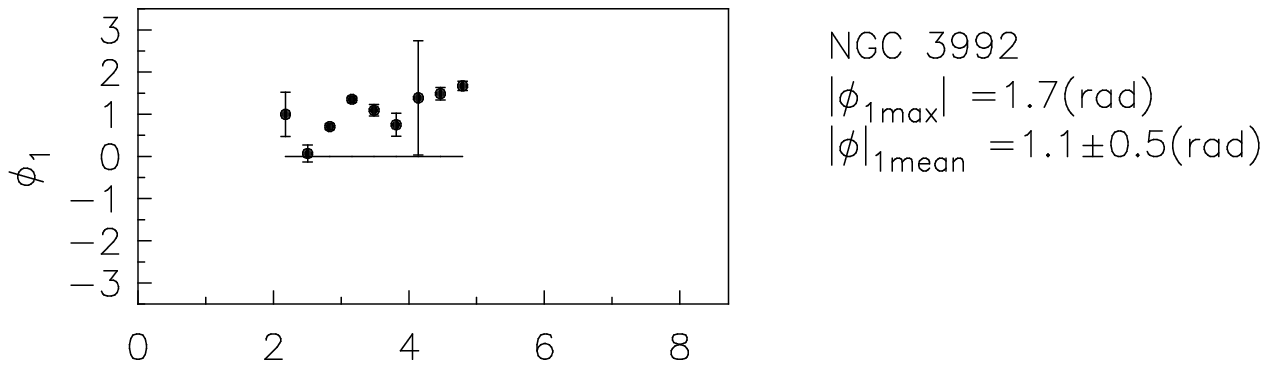}
\includegraphics[width=84mm,height=25mm]{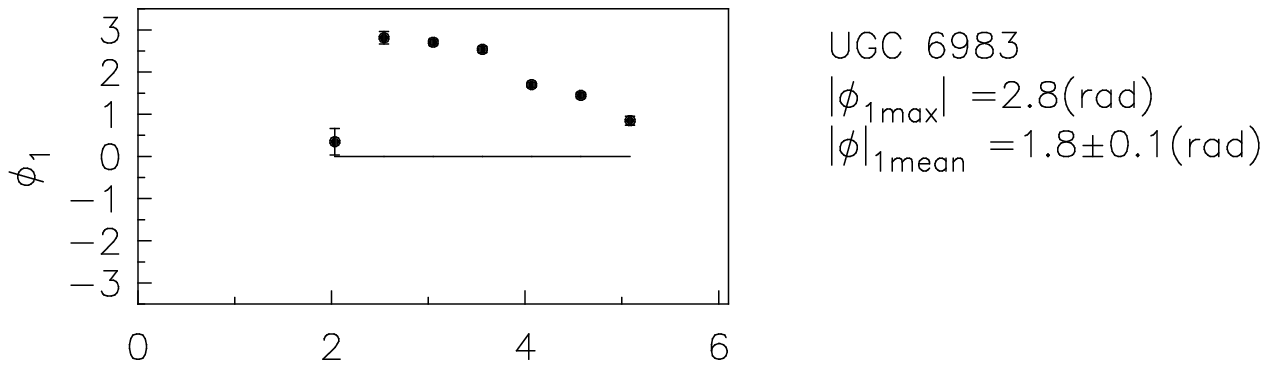}
\includegraphics[width=84mm,height=25mm]{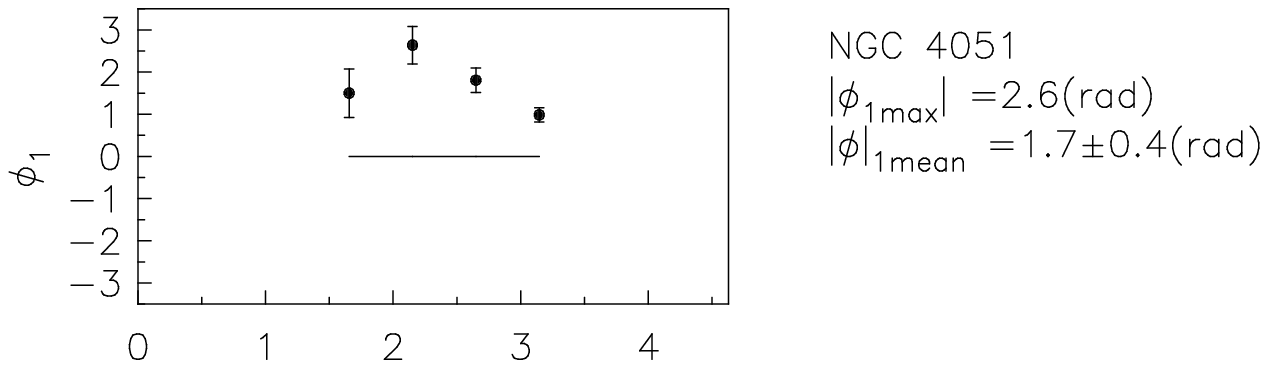}
\includegraphics[width=84mm,height=25mm]{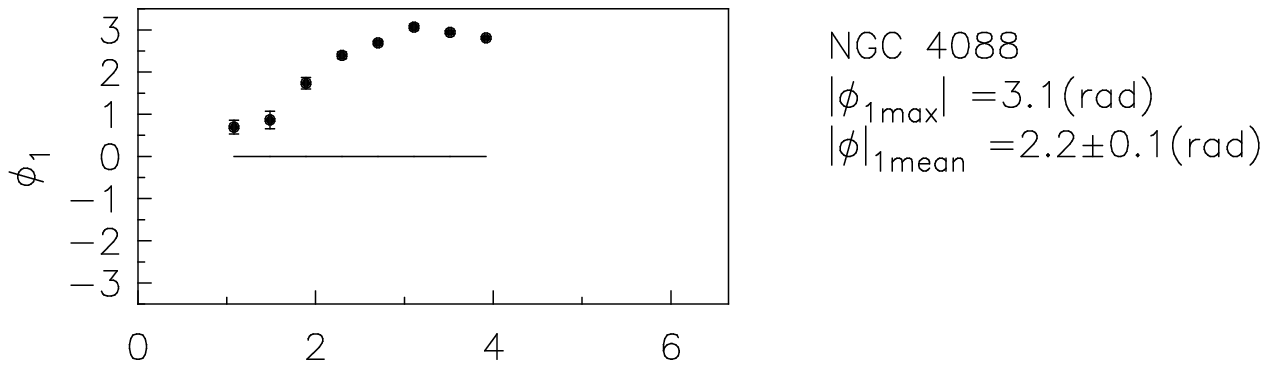}
\includegraphics[width=84mm,height=25mm]{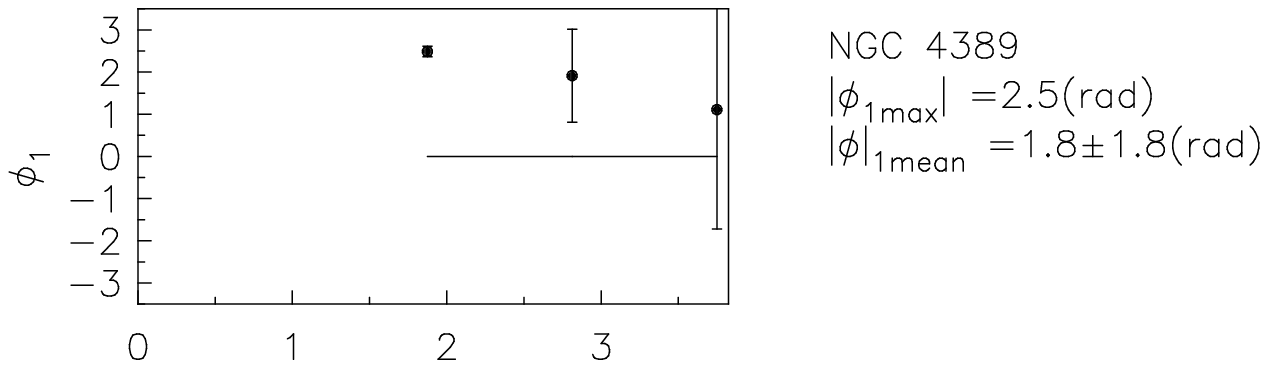}
\caption{ The asymmetry phase parameter derived from the surface density maps (moment 0). In each of the maps, the radius is in the units of K'-band scale length.}

\end{figure*}

The results from Figs. 1-2 and Table 2 can be summarised as:
\begin{enumerate}
\item{The amplitude of disk lopsidedness shows an increase with
radius, and the phase is nearly constant with radius
as also seen earlier for the field galaxies (Rix \& Zaritsky
1995), and the group galaxies (Paper I). We note, however, that these correlations are not so robust in the 
Ursa Major case. These similarities, and the differences, 
provide important clues to the origin of the lopsidedness in the field and group galaxies.

In contrast, the advanced mergers of galaxies show an amplitude $A_1$ 
that peaks at an intermediate radius of a few kpc and then turns over, 
and the phase
shows a large fluctuation with radius (Jog \& Maybhate 2006).
These two different properties  clearly 
underline the different mechanism for the origin of lopsidedness
in the present sample of normal galaxies 
as compared to the mergers of  galaxies.}

\item{The average value of the lopsidedness is 
$\sim 0.14 \pm 0.05$ (Table 2,column 4). This is similar to the mean
 value for the
field galaxies obtained over the same radial region of
1.5-2.5 disk scalelengths  (Rix \& Zaritsky 1995, Bournaud et
al. 2005) and about half of the average value that is seen in the Eridanus group (Paper I). In addition, only 2 out of the 11 galaxies (or $\sim 20\%$) of this sample have A$_{1}\ge 0.2$ and none have A$_{1}\ge 0.3$. On the other hand, the Eridanus group of galaxies (Paper I)   showed higher values ($\sim 40\%$ and $\sim 30\%$ of the sample had A$_{1}$ values higher than 0.2 and 0.3 respectively)}.

Thus, despite being in a group environment, the Ursa Major galaxies
show overall smaller $A_1$ values; this point is highlighted in
 Figure 3 where histograms for the $A_1$ values for the Ursa
Major and Eridanus groups are  plotted. To verify that this is not a spurious effect due to the limited size of the samples, we have carried out Kolmogorov-Smirnov (KS) test on the data samples, including 3 values of A$_1$ estimated from R-Band analysis (see Discussion ). The D statistics value was estimated to be $0.357$. The probability that the two samples come from the same distribution is 23.6\%.
This points to a different physical reason for the origin of the observed lopsidedness in this group (see Section 4 for a detailed
discussion).

\item{In addition, since the HI extends much farther out and can be
studied to a larger radii than the stars, we have measured $A_1$
values to larger radial range going up to 4-6 disc scalelengths as compared to
2.5 disc scalelengths possible in the stellar case (Rix \& Zaritsky 1995,
also see Section 3.2 in the present paper).}

\item{The values of asymmetry as measured by the Fourier amplitudes $A_1$, $A_2$, 
and $A_3$ over the range 1-2 $R_w$ (see Table 2) are comparable. In contrast, 
the field galaxies show the amplitudes $A_1$ and A$_2$ for m=1,2 to be stronger than 
$A_3$ for m=3 in general (Rix \& Zaritsky 1995, Bournaud et al. 2005), and in centres of advanced mergers it was found that $A_3$ is large only when $A_1$ is large, and in any case the $A_3$ values are always smaller than the $A_1$ values (Jog \& Maybhate 2006).
The similar values of the amplitudes of m=1,2,3 in the Ursa Major case could be due to the complex group potential with possible  multiple interactions which may be reflected 
in the amplitudes of m=3 and higher modes.}

\end{enumerate}
\begin{figure*}
\includegraphics{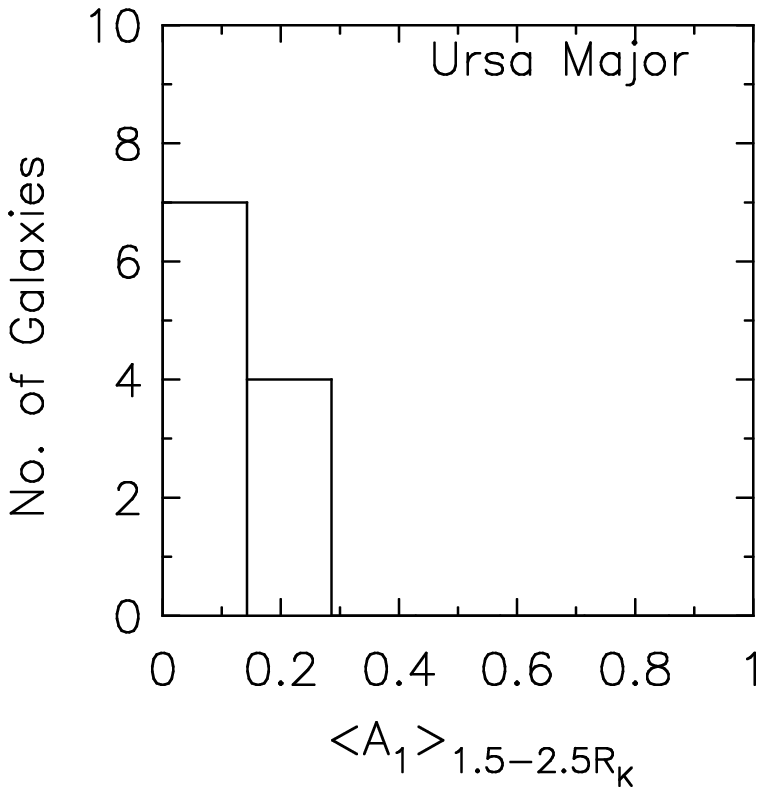}
\includegraphics{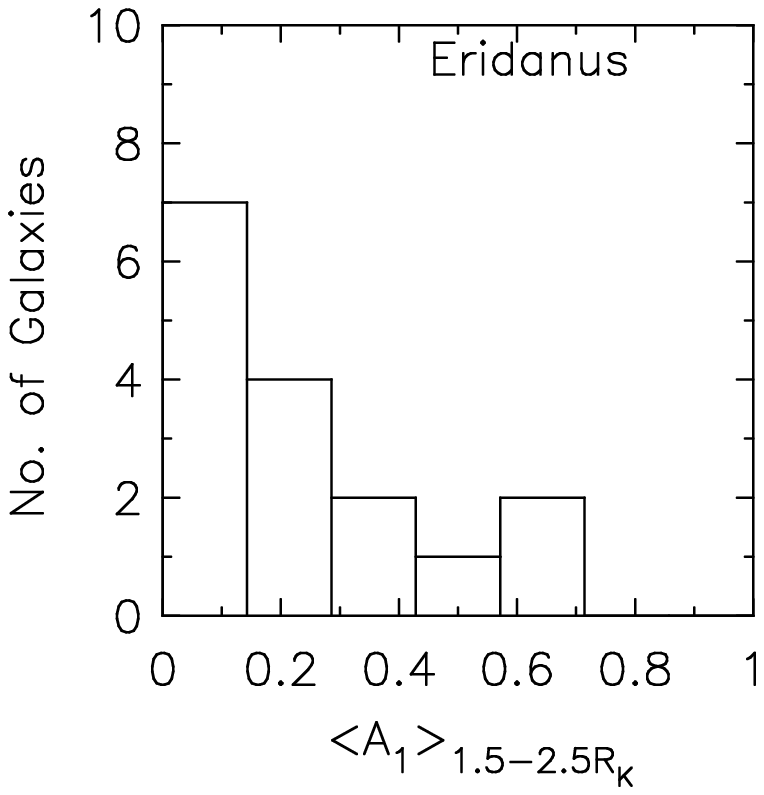}
\caption{ The  histograms showing the  number of galaxies vs.
$<A_1>$ in the 1.5 to 2.5 $R_{K'}$ range for the Ursa Major Group (left) and the Eridanus Group (right) of galaxies.
Clearly, the Ursa Major galaxies show overall smaller amplitudes of lopsidedness.}

\end{figure*}

\begin{table*}
\centering
\noindent
\caption{ The mean values of $A_1$ in the range 1.5-2.5 $R_{K'}$, and $A_1$, $A_2$, $A_3$ in the range 1-2 $R_w$ }
\begin{tabular}{@{}lccccccc@{}}
\hline 
\hline 
\bf{Name}& Hubble Type&$R_{K'}$&$<A_1>_{K'}$&$R_w$&$<A_1>_w$&$<A_2>_w$&$<A_3>_w$\\
	 &	       &(kpc) &$1.5-2.5R_{K'}$&(kpc)&$1-2 R_w$&$1-2 R_w$&$1-2 R_w$\\
\hline
UGC 6446 &7 &0.82 &0.14&3.19 &0.17&0.23& 0.08 \\
NGC 3726 &5 &2.12 &0.11&4.08 &0.16&0.12&0.12\\
NGC 3893 &5 &1.70 &0.20&$--$ &$--$&$--$&$--$ \\
NGC 3949 &4 &1.00 &0.16&2.69 &0.22&0.27&0.07 \\
NGC 3953 &4 &2.90 &0.13&3.96 &0.17&0.28&0.17 \\
UGC 6917 &7 &1.90 &0.13&3.80 &0.13&0.21&0.06 \\
NGC 3992 &4 &3.11 &0.23&$--$ &$--$&$--$&$--$\\
UGC 6983 &6 &2.06 &0.03&4.43 &0.13&0.07&0.11 \\
NGC 4051 &4 &1.37 &0.15&2.86 &0.17&0.25&0.14 \\
NGC 4088 &4 &1.81 &0.08&4.13 &0.19&0.18&0.11 \\
NGC 4389 &4 &0.74 &0.14&$--$ &$--$&$--$&$--$ \\
\hline
Mean	&  &	  &$0.14\pm0.05$    &     &$0.17\pm0.03$&$0.20\pm0.07$&$0.11\pm0.04$ \\
\hline
\hline\\
\end{tabular}
\end{table*}

\subsubsection{Kinematical Lopsidedness in HI}

The five parameters i.e. the coordinates of the centre ($x_{0},y_{0}$), systemic velocity ($V_{0}$), circular velocity ($V_{c}$), inclination ($i$) and the position angle ($PA$),  estimated from the velocity maps using the iterative use of GIPSY routine ROTCUR. These parameters were given as the input to another GIPSY routine called RESWRI along
with the velocity maps and the HI-surface density maps to obtain the harmonic coefficients.

At each radii (r), the line of sight velocity was expanded in the form

\begin{equation}
v_{los}(r,\p1)=c_0+\sum_{m=1}c_m\cos(m\p1)+s_m\sin(m\p1)
\end{equation}

where, $c_m$,$s_m$ are the harmonic coefficients, $c_0$ is identical to the systemic velocity $V_{0}$ and $\p1$ is the azimuthal angle. These
harmonic coefficients were derived at concentric radii which were separated by $15^{\prime\prime}$.In our analysis we have derived the harmonic coefficients up to the $10^{th}$ order. This was partially prompted by the observation that effects of  bars tend to retain the strength of the Fourier coefficients even for m=10 terms \citep{Buta03}. Typical velocity harmonic
coefficients are shown in Figure 4.

\begin{figure*}
\includegraphics[width=185mm,height=165mm]{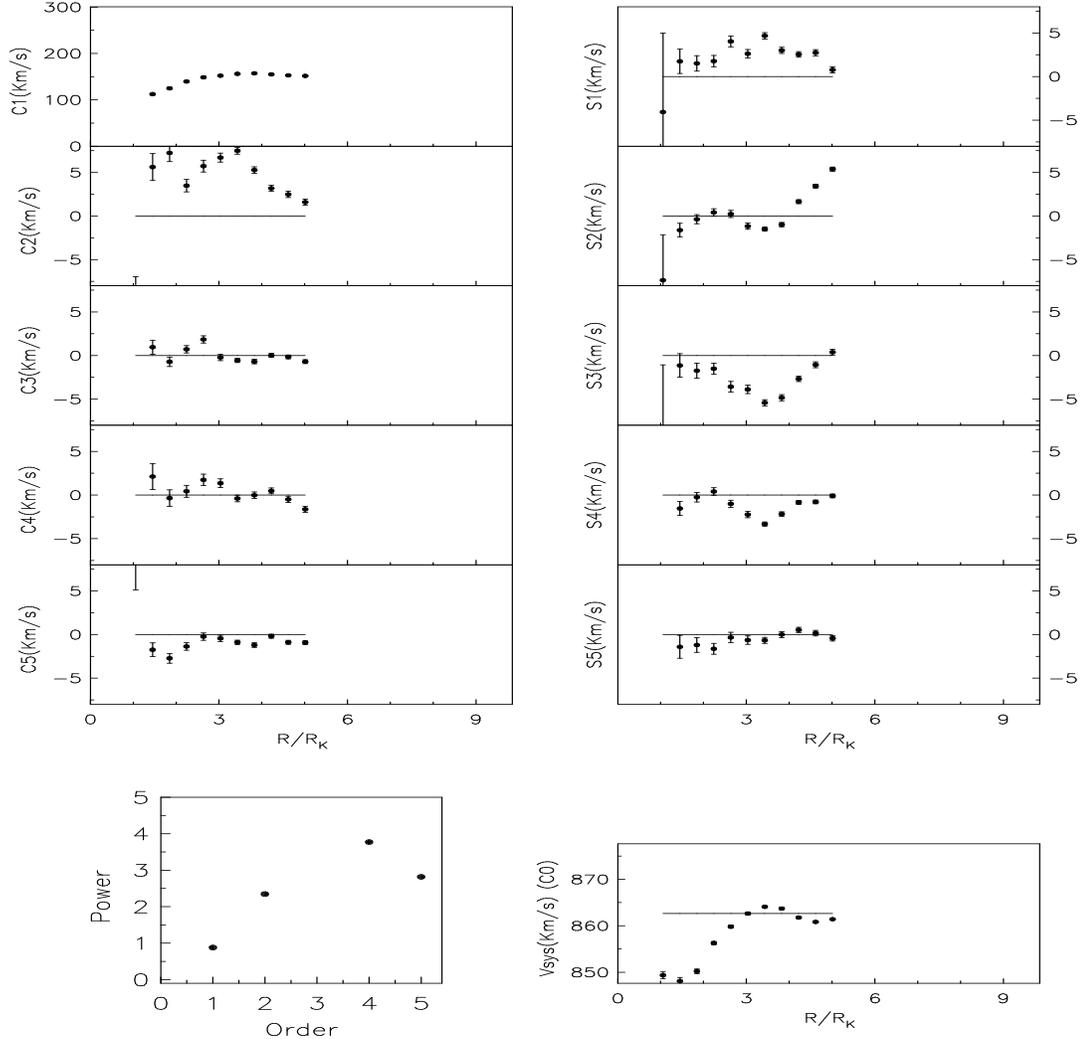}

\caption{Velocity harmonic coefficients estimated for NGC 3726. The power in each of the harmonic order is shown in the bottom left hand panel. The bottom right hand panel shows the $c_0$ coefficient ($V_{sys}$) as a function of radius.}

\end{figure*}

From these coefficients, since $c_3 \sim 0$, it is seen that the 
inclination fitting has converged \citep{Schoen97}. In addition, an estimate of
the effects of spiral arms and global elongation in the potential 
that gives rise to the kinematical lopsidedness can be obtained from 
the $s_1$ \& $s_3$ coefficients \citep{Schoen97}. If the influence 
of spiral arms are large, $s_1$ and $s_3$ are expected to
oscillate rapidly \citep{Schoen97}. Since this is not seen in Figure 4, their contribution
must be small. 
 
From the harmonic coefficients thus obtained, it is possible to estimate the 
elongation in the potential of the galaxy times
 a factor of $\sin(2\phi_2)$, where $\phi_2$ is the 
phase angle for $m=2$ in the plane of the galaxy \citep{Schoen97}. 
The estimation of the ellipticity of the potential of an early type galaxy, 
IC 2006, using the velocity field for the HI ring in it was first
carried out by Franx et al. (1994). This was later generalised to include 
spiral galaxies with extended exponential disks \citep{Schoen97}. In this
procedure, with the assumptions of flat rotation curve in the outer regions of the galaxy and constant
phase ($\phi_2 (r) $), the ellipticity of potential ($\epsilon_{pot}$) is obtained as
$\epsilon_{2}\sin(2\phi_{2}(r))$ \citep{Schoen97}.
We have carried out similar analysis for all 
the sample galaxies in the Ursa Major group, and presented
in Section 3.3.
A typical asymmetry or the elongation in the potential for NGC 3726 is shown 
in Figure 5.

\begin{figure*}
\includegraphics[width=110mm,height=70mm]{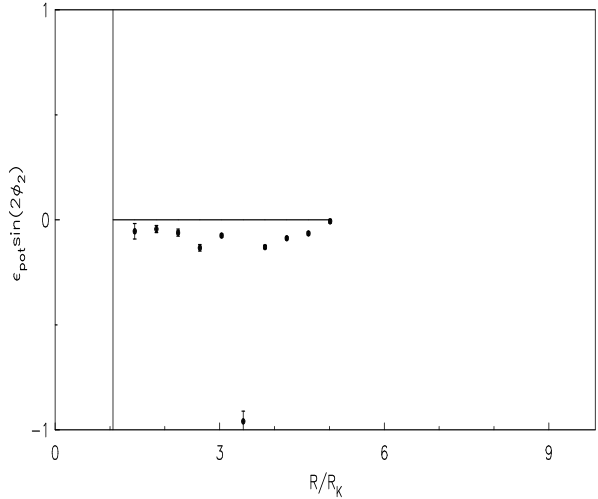}
\caption{ The estimated elongation in potential $\epsilon_{pot}
= \epsilon_2\sin(2\phi)$ derived from 
the velocity harmonic coefficients for NGC 3726}
\end{figure*}

\subsection{Harmonic Analysis of Optical and Near-IR Data}
The harmonic analysis of the optical (R-band and K'-band) data was carried out as per 
the procedure adopted by 
earlier workers \citep{Rix95, Zaritsky97, Angiras06}. 
The original images obtained from CADC were corrected for the sky background 
and for the atmospheric extinction. In addition to this, foreground stars were masked.
The optical centres of the galaxies were estimated using the IRAF
task IMCNTR. It was seen that the optical centres of these galaxies were the same as that obtained by Tully et al. (1996). These images were
deprojected using the IRAF {\footnote {IRAF is distributed by the National Optical Astronomy Observatories,
which are operated by the Association of Universities for Research in Astronomy, Inc., under cooperative agreement with
the National Science Foundation.}} task IMLINTRAN \citep{Buta98}. In this deprojection, we have not taken into account
the effects of the bulge of the galaxy. The effect of bulge is expected to be very small as we are mainly interested in
the outer regions of the galaxy \citep{Bournaud05}. In addition we expect bulge contamination to be serious for the m=2 mode and for almost edge-on galaxies which is not relevant in our case. For each of the galaxies, along various concentric annuli, the surface density as a function of angle was
extracted using the ELLIPSE {\footnote {ELLIPSE is a product of the Space Telescope Science Institute, which is
operated by AURA for NASA.}}task. Each of the rings were separated by $1^{\prime\prime}$ (typical resolution) in the case of R-Band images. Harmonic analysis was carried out on the extracted
surface density values and normalised coefficients were estimated. The variation of $A_1$ coefficients,
derived from R-Band image along with those derived from the HI analysis for 
two sample galaxies NGC 3726, NGC 4051 are shown 
in Figure 6.

\begin{figure*}
\includegraphics[width=84mm,angle=-90]{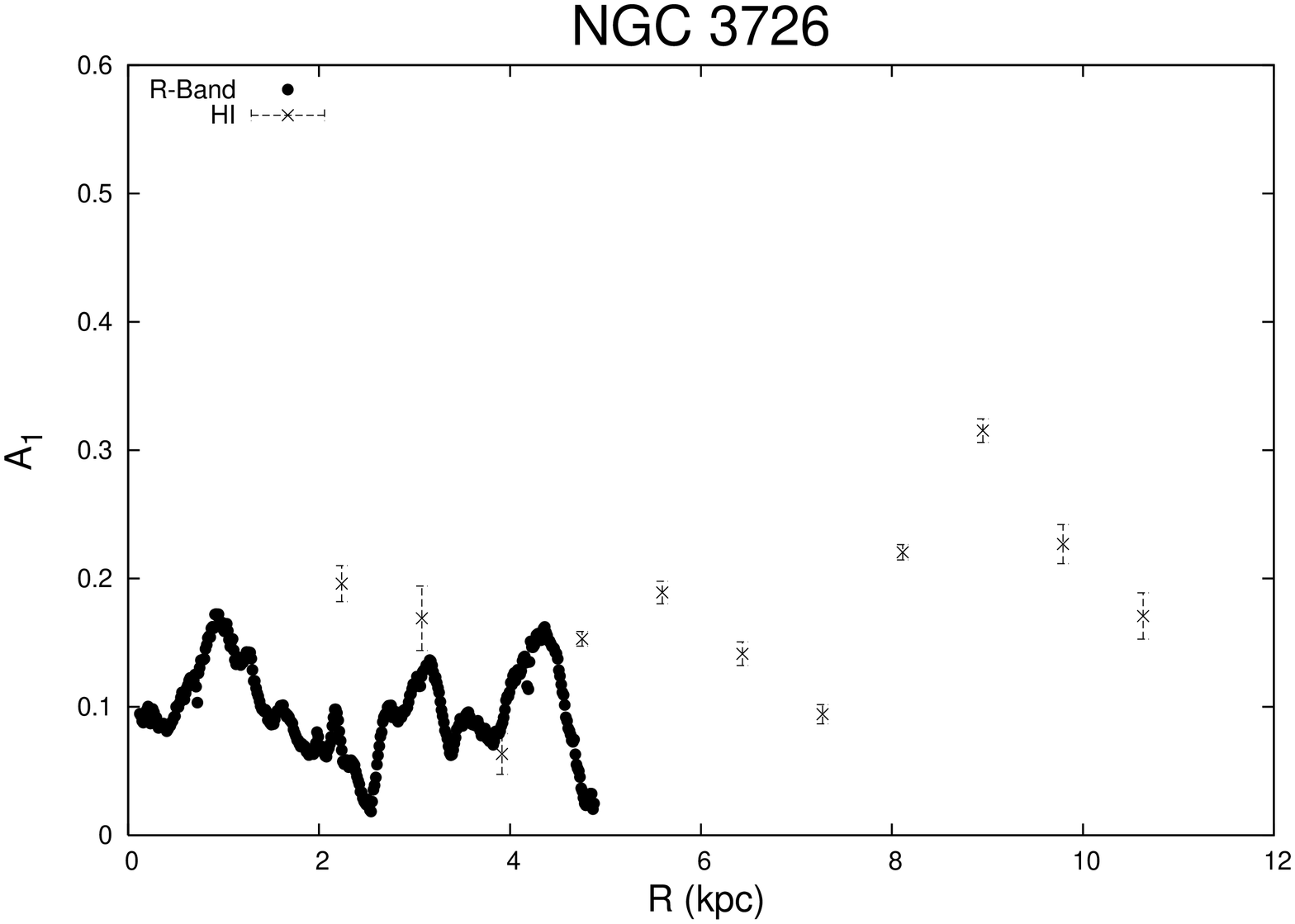}
\includegraphics[width=84mm,angle=-90]{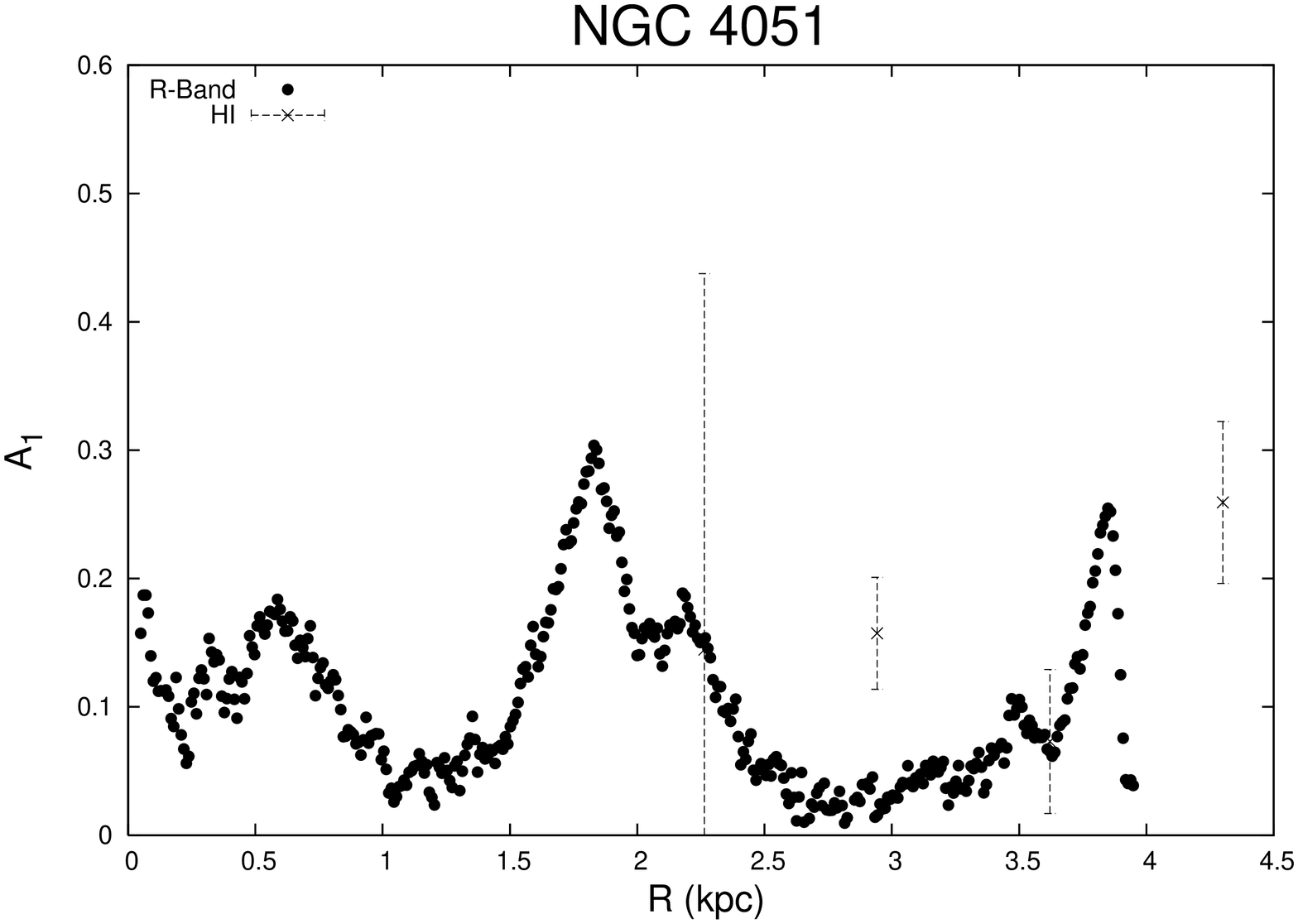}
\caption{The $A_1$ coefficients derived from R-Band images of galaxies, 
which are compared
with that obtained from HI surface density maps. Note that the values are comparable in the inner regions 
of radial overlap.}

\end{figure*}

From this figure, it can be seen that the lopsidedness in the
stellar distribution measured 
from the optical data is comparable to that
seen in HI in the same radial region of study. 
Thus these values represent true lopsidedness and not features seen only for the
kind of tracer used. This agreement confirms what
was also seen in our earlier study of HI lopsidedness in
Eridanus (Paper I), and provides evidence that the
asymmetry in both arise due to the stars and gas responding respectively to
the same perturbation potential as proposed by Jog (1997). 
In the outer parts, only HI is available as a tracer 
since the near-IR data is available only  up to 2.5 disc 
scalelengths (see Section 3.1.1).

\subsection{Strength of the Perturbation Potential}

Assuming that the asymmetry arises due to the disc response to a
distorted halo (Jog 1997), from the surface density maps, we can obtain the
perturbation potential corresponding to the m=1,2,3 terms from
the observed amplitudes $A_1$, $A_2$ and $A_3$ for the normalised Fourier coefficients (Jog 2000). Here the perturbation potential is taken to be of
the form ${V_c}^2 \epsilon_m  cos m\phi $ where $V_c$ is the
flat rotation curve value and $\epsilon_m$ denotes the
perturbation parameter. As a result, it can be shown that $\epsilon_1$ denotes the lopsided potential,
and $\epsilon_2$ denotes the elongation or ellipticity of the
perturbation potential.

The observed HI distribution was fitted with a Gaussian curve and the associated {\it Gaussian scalelength} $R_w$ was estimated. Using this scalelength the relations between $\epsilon_m$, $A_m$ and $R_w$ have been derived for m = 1, 2 and 3, following the procedure in Jog (2000) where it was developed for an exponential disk distribution
in a region of flat rotation curve. It is now believed that haloes of galaxies in groups are merged in the outer parts \citep{Athanassoula97}, however in the inner regions of 10 kpc where we study the asymmetry, we can still treat the asymmetry as if the halo were isolated. Hence the above model of disk response to a halo is still reasonably valid and the inner regions could carry the signatures of complex tidal interactions in a group setting.
This yields the following relations between the perturbation
parameters for the potential $\epsilon_m$ and the $A_m$ values:

\begin{equation}
\epsilon_2 = \frac {A_2(r)} {(r/R_w)^2 + 1} 
\end{equation}

and

\begin{equation}
\epsilon_3 = \frac {A_3(r)} {(2/7){(r/R_w)^2} + 1}
\end{equation}

The relation for m=1 was already obtained in Paper I,
and is:

\begin{equation}
\epsilon_1 = \frac {A_1(r)}  {2(r/R_w)^2 - 1}
\end{equation}

The resulting mean values of the perturbation parameters are obtained using the measured values of the Fourier amplitudes $A_1$, $A_2$ and $A_3$, and are given in Table 3.

Using the analysis of the kinematical data, we have also obtained
$\epsilon_{2} \sin [2 \phi_2 (r)]$ for the galaxies in Ursa Major
as discussed in Section 3.1.1. The mean values of this quantity in the range 1-2 $R_w$ are shown in 
Table 3.

The main results from this subsection are:
\begin{enumerate}
\item{The average value of $\epsilon_1$ or the lopsided
perturbation for the
potential  obtained in the outer parts (in the radial range 1-2 $R_w$) is 
$\sim 6 \%$, this is smaller than the value for  
the Eridanus case where the halo lopsided potential was derived to be $10 \%$ - this reflects the smaller observed amplitudes of lopsidedness in the present sample.

Thus if the lopsidedness is due to the response to the halo
distortion, then this gives 6\% as the typical halo
lopsidedness for the Ursa Major galaxies.}

\item{The elongation in the potential or the magnitude or the amplitude of the elongation in the term $\epsilon_2$ value is comparable (within a factor of $\sin(2\phi_2)$)
whether calculated from the observed spatial asymmetry or from the kinematical asymmetry (see columns 4 and 6 of Table 3). This confirms the argument (Jog 1997, Jog 2002) that both spatial and kinematical asymmetry
result from the same perturbation potential.}

\item{The values of all three perturbation potentials derived $\epsilon_1, \epsilon_2 , \epsilon_3$  are
comparable. Although this result depends on the model used, it reinforces the similar result obtained for the Fourier amplitudes which are directly observed and hence are model-independent (Section 3.1.1).  This can be an important clue to the mechanism for
generating lopsidedness in groups, and perhaps indicates the importance of  multiple simultaneous tidal interactions  that can occur under the special conditions of a group environment.}
\end{enumerate}

\begin{table*}
\centering
\noindent
\caption{The HI-scalelength, and the mean perturbation parameters of potentials obtained from $A_1$, $A_2$ and $A_3$- the coefficients of surface densities and from velocity fields. The mean values are calculated between 1-2 $R_w$ }
\begin{tabular}{@{}lccccc@{}}
\hline
\hline
\bf{Name}&HI Scalelength ($R_w$) (kpc)&$<\epsilon_{1}>$&$<\epsilon_{2}>$&$<\epsilon_{3}>$&$<\epsilon_{2}\sin(2\phi_2)>$\\
\hline
UGC 6446       &3.19     &0.046      &0.040     & 0.065     &-0.173\\
NGC 3726       &4.08     &0.049      &0.042     & 0.111     &-0.392\\
NGC 3949       &2.69     &0.072      &0.106     & 0.056     & 0.178\\
NGC 3953       &3.96     &0.044      &0.098     & 0.113     &-0.171\\
UGC 6917       &3.80     &0.041      &0.057     & 0.083     & 0.007\\
UGC 6983       &4.43     &0.032      &0.022     & 0.128     &-0.213\\
NGC 4051       &2.86     &0.082      &0.113     & 0.109     &-0.087\\
NGC 4088       &4.13     &0.089      &0.076     & 0.092     & 0.001\\
\hline\\
Mean	       &	 &$0.057\pm0.021$&$0.069\pm0.034$&$0.095\pm0.025$&$-0.106\pm0.172$\\
\hline
\hline\\
\end{tabular}
\end{table*}

\section{Discussion : Lopsidedness in groups}
\begin{enumerate}
\item {The Ursa Major group of galaxies show a typical lopsidedness of $\sim
14 \% $ in the
inner regions, that is comparable to the field case, and about
half of what is seen in the Eridanus group (see Section 3.1.1
for details).

We also measure the $A_1$ values in the outer parts between 1-2
$R_w$, and find this to be $\sim 17 \%$. Again this is smaller by a
factor of $\sim 1.6$ compared to the Eridanus case. }

\item{We plot  the mean $A_1$ ($<A_1>$) value in the inner regions of the 
galaxies with respect to Hubble type in Figure 7, where we also
plot the corresponding values from the Eridanus study for comparison (Paper I, Figure 5) .
Note that the early-type galaxies in the Eridanus group show a higher lopsidedness, this
is opposite to what is seen in the field galaxies \citep{Zaritsky97,Bournaud05}, and points to tidal interactions as the mechanism for the origin of lopsidedness as argued in Paper I. In contrast, in field galaxies, gas accretion plays an important role in generating lopsidedness \citep{Bournaud05}. 
The anti-correlation with galaxy Hubble type is weaker in the
Ursa group case, perhaps because our sample here only covers a smaller
subgroup of galaxy types from type 4 to 7, whereas the Eridanus
study spans a much larger range from type 1 to 9.

To address this issue, we obtained the R-Band images of three more galaxies from the Ursa Major sample \citep{Tully96}, namely UGC 6930, NGC 4102 and NGC 3729, from the Canadian Astronomy Data Centre (CADC) on which Fourier analysis was carried out. UGC 6930 belonged to Hubble type 7 and had an inclination of 32$^\circ$. NGC 4102 and NGC 3729 belonged to Hubble type 2 and had inclinations 58$^\circ$ and 48$^\circ$ respectively. The A$_1$ coefficient in the range 1.5 to 2.5 $R_{K'}$ for UGC 6930 was estimated to be 0.02. In the same range, the A$_1$ coefficients for NGC 4102 and NGC3729 were 0.12 and 0.04 respectively. These galaxies were not included in our HI analysis because of lack of reliable HI data. It should be noted that if these three points are included in KS-test, the probability that the values of A$_1$ come from the same distribution is $23.6\%$ while the maximum difference (D) between the cumulative distribution is 0.357.    
 The three resulting points are shown in Fig. 7 (denoted by symbol \otriangle) and they confirm that
 the distribution of A$_1$ vs. R is nearly flat. Thus there is no clear anti-correlation in this case unlike that seen in the Eridanus group. However, this distribution does not show a positive correlation with the Hubble type either, unlike the field case \citep{Bournaud05}.} Thus Figure 7 confirms the different physical origins for the lopsidedness in the group and field cases. It also confirms that the anti-correlation seen in the Eridanus group can be attributed to the group environment, and requires a higher galaxy number density as seen in the Eridanus to be effective.

\item {The kinematical analysis gives a value for the elongation in
the potential, showing all galaxies where such analysis could be
carried out to be disturbed. This was found earlier for a group
of five galaxies in the Sculptor group galaxies (Schoenmakers 2000).}

\item {The above results show that the group environment is
conducive to producing lopsidedness, with tidal interactions playing a major role in this. Galaxy interactions can give rise both to lopsidedness and a secular evolution towards
early-type galaxies as argued by Bournaud et al. (2005).
There are indeed some indications of tidal interactions in Ursa Major 
group of galaxies (Verheijen \& Sancisi 2001).}
\end{enumerate}

\begin{figure*}
\includegraphics[width=84mm,angle=-90]{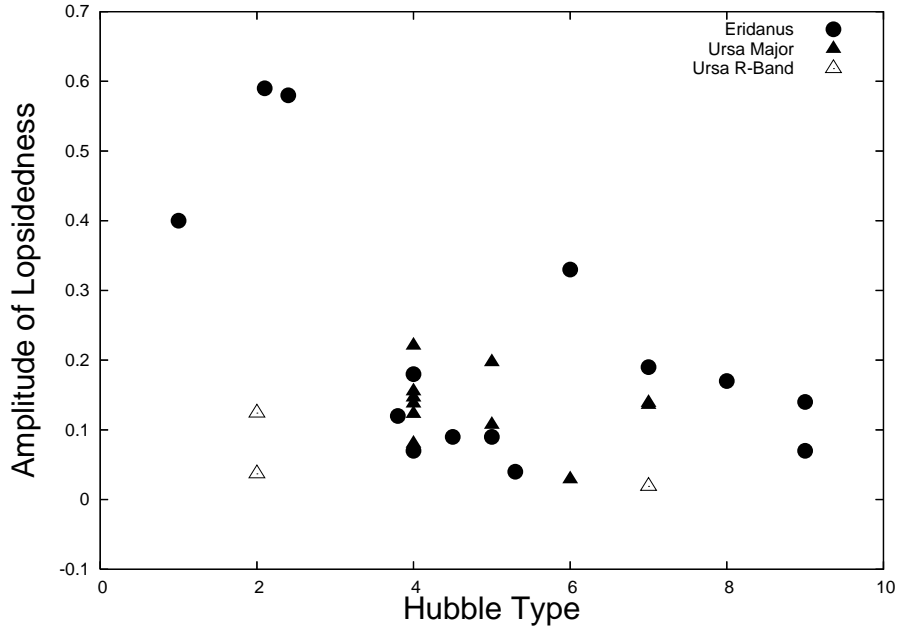}
\caption{The $<A_1>$ values in the 1.5 to 2.5 K$^\prime$-band scale length of Ursa Major galaxies. The Eridanus Group
values are taken from Paper I.  The values denoted by \otriangle, correspond to A$_1$ values of the 3 galaxies estimated from R-Band analysis (see Discussion ).The A$_1$ values are higher for the early-type galaxies in the Eridanus while the distribution is flatter for the Ursa Major group.}
\end{figure*}

The Ursa Major group is a loose group, and has number density that is intermediate between the field and the Eridanus values. Hence tidal interactions are less important in the Ursa Major group, which could explain the lower amplitude of lopsidedness (A$_1$).

The lower values of asymmetry parameter ($<A_1>$) observed for Ursa Major group of galaxies may also find a partial
explanation, if we assume that it mainly falls on a filament and is in the process of forming a group. Such a process
is observed in ZwCl 2341.1+0000 \citep{Bagchi02}.

\section{Conclusions}

The main conclusions drawn from this paper are as follows:
\begin{enumerate}
\item{The mean amplitude of disk lopsidedness in the Ursa Major group galaxies is 
measured and found to be comparable to the field sample, while the 
Eridanus group showed a factor of two higher lopsidedness. 

The smaller
amplitudes of lopsidedness seen in the present study
could be due to the lower galaxy number density and the lower velocity
dispersion in the Ursa Major group (see Section 2). The group environment and tidal interactions are shown to play a major role in generating lopsidedness, especially in a denser group like the Eridanus. 

The disk lopsidedness can thus be used as diagnostics to study the
galaxy interactions and the halo properties in groups of
galaxies.}

\item{The values of elongation of potential as measured from
spatial and kinematical studies gives comparable values, thus
supporting the idea  (Jog 1997) that both types of asymmetry arise due to
the same perturbation potential.}

\item{The values of the asymmetry as measured by the mean fractional Fourier amplitudes A$_1$, A$_2$ and A$_3$ are found to be comparable,
and also the derived perturbation potential parameters $\epsilon_1$, $\epsilon_2$ and $\epsilon_3$ are found to be comparable. This is in
contrast to the field galaxies where A$_1$ and A$_2$ are stronger than A$_3$ and higher mode amplitudes. This indicates the importance of multiple tidal interactions that can occur under the special conditions of a group environment.}
\end{enumerate}

\section{ACKNOWLEDGMENTS}

We thank the referee, Frederic Bournaud, for a careful reading of the manuscript 
and for the critical comments and the suggestion of including the A$_1$ values from R-Band images in Figure 7. These have improved the presentation of the paper.
RAA takes great pleasure in thanking K. Indulekha, School of Pure and Applied Physics, M.G.University,for her constant encouragement during this project. He also thanks the University Grants Commission of India and St.Joseph's College, Bangalore for granting study leave under the FIP leave of 10th five year plan and Raman Research Institute, Bangalore for providing all the facilities to pursue this study.This research used the facilities of the Canadian Astronomy Data Centre operated by the National Research Council of Canada with the support of the Canadian Space Agency.

\end{document}